\documentclass[twocolumn,showpacs,preprintnumbers,amsmath,amssymb]{revtex4}
\usepackage{graphicx}

\DeclareFontFamily{OT1}{pzc}{}
\DeclareFontShape{OT1}{pzc}{m}{it}{<-> s * [1.10] pzcmi7t}{}
\DeclareMathAlphabet{\mathpzc}{OT1}{pzc}{m}{it}

\begin{document}

\title{Relaxation and curvature-induced molecular flows within multicomponent 
membranes}

\author{Richard G.~Morris}

\affiliation{Theoretical Physics, The University of Warwick, Coventry, CV4 7AL, 
UK.}

\email{r.g.morris@warwick.ac.uk}

\begin{abstract}
	The quantitative understanding of membranes is still rooted in
	work performed in the 1970s by Helfrich and others, concerning amphiphilic
	bilayers.  However, most biological membranes contain a wide variety of
	nonamphiphilic molecules too.  Drawing analogy with the physics of
	nematic/non-nematic mixtures, we present a dynamical (out-of-equilibrium)
	description of such multicomponent membranes.  The approach combines
	nematohydrodynamics in the linear regime and a proper use of
	(differential-) geometry.  The main result is to demonstrate that one can
	obtain equations describing a cross-diffusion effect (similar to the Soret
	and Dufour effects) between curvature and the (in-membrane) flow of 
	amphiphilic molecules relative to nonamphiphilic ones.  Surprisingly, the 
	shape of a membrane relaxes according to a simple heat equation in the mean 
	curvature, a process that is accompanied by a simultaneous boost to the 
	diffusion of amphiphiles away from regions of high curvature.  The model 
	also predicts the inverse process, by which the forced bending of a 
	membrane induces a flow of amphiphilic molecules towards areas of high 
	curvature.  In principle, numerical values for the relevant diffusion 
	coefficients should be verifiable by experiment.
\end{abstract}


\maketitle
\section{Introduction}
The behaviour of the cell membrane is crucially important to a wide variety of 
processes in biology~\cite{BA+89}.  However, since the underlying construction 
of almost all biological membranes is that of an amphiphilic bilayer, much of 
the physics literature has so far focused on understanding simple bilayers and 
their closed-form counterparts, vesicles~\cite{DDL+95,US96}.  Despite intensive 
research over the last 50 years, such approaches have failed to properly 
describe the role of the biological membranes in commonly observed phenomena, 
such as cell locomotion for example.  The point is that often the 
nonamphiphilic component of the membrane is important (in the case of cell 
locomotion, consider so-called membrane-to-cortex attachment proteins embedded 
in the bilayer).  In this paper we attempt a consistent dynamical treatment of 
such multicomponent membranes, which makes quantitative predictions about the 
dynamical relationship between curvature and the diffusion of amphiphiles 
relative to non-amphiphiles.  The hope is that such ideas might act as a base 
onto which successful predictive models of the above phenomena can be built, 
and more.

The modern quantitative description of bilayers was pioneered by 
Helfrich~\cite{WH73} in the 1970s (and independently by Canham~\cite{PBC70} and 
Evans~\cite{EAE74}) and has not advanced greatly since.  Drawing parallels 
between the amphiphiles of a bilayer and the rodlike molecules of nematic 
liquid crystals, Helfrich adapted expressions for the Frank free 
energy~\cite{FCF58} by replacing the director with the normal to the membrane 
surface.  The result was a local free energy per unit area of the form
\begin{equation}
	\frac{\kappa}{2}\left( 2H - C_0 \right)^2 + \kappa_g K.
	\label{eq:Helfrich}
\end{equation}
In the normal way, $H$ and $K$ are just the mean and Gaussian curvatures 
respectively~\cite{SchaumTenCalc}, and $C_0$--- the spontaneous curvature--- is 
a constant.  The quantities $\kappa$ and $\kappa_g$ are called the bending 
rigidity and elastic modulus of Gaussian curvature, respectively, and are also
constant.

Strictly, Eq.~(\ref{eq:Helfrich}) describes the free energy of a monolayer 
rather than a bilayer, and therefore a number of attempts at improvement have 
been made (\textit{e.g.},~\cite{SS+85,LM+94}).  However, due to the similarity 
between monolayers with different head-tail interactions and bilayers with 
differing leaf densities, both the original model and its relatives have been 
used extensively in the physics literature~\cite{US96}, most notably in the 
static description of vesicles under 
constraint~\cite{HJD+76,JTJ77,JCL+79,JCL82,MAP85,USKB+91,H-GD+97}.  Only 
recently have attempts been made to write nonequilibrium descriptions of 
membrane dynamics, but so far the focus has been solely on 
vesicles~\cite{DFAJM08,BBSS09,DFAJM09,RGM+10,RMAJM11}.  Taking inspiration from 
such studies, this article goes back to the original model of Helfrich and 
demonstrates that it can be extended by analogy with nematohydrodynamics in the 
linear regime.  As we show, such an approach permits the incorporation of 
additional nonamphiphilic components and leads to a membrane description in 
terms of both curvature and in-membrane molecular flows.

\section{Free energy of the membrane}
The starting point is to make two important observations.  First, biological 
membranes are made not only from amphiphiles, but from a whole host of other 
nonamphiphilic molecules (cholesterol, carbohydrates, proteins, protein 
channels, fat-soluble molecules \textit{etc.}).  Second, in most biological 
systems the temperature is sufficiently high that the molecules in the membrane 
effectively form a two-dimensional fluid which, when close to equilibrium, can 
be described by linear nonequilibrium thermodynamics 
(LNET)~\cite{NonEquilThermDeGrootMazur}.  In this framework, the geometry of 
the problem then enters through the Gibbs free energy, so it is important to 
ensure a consistent thermodynamic formulation.  Recall then that LNET assumes a 
\textit{local} equilibrium, such that the Gibbs relation
\begin{equation}
	T\mathrm{d}s = \mathrm{d}u + p\mathrm{d}\nu - 
	\left\{\mathrm{d}g\right\}_{T,\ p},
	\label{eq:Gibbs}
\end{equation}
is obeyed at every point in space and time.  Here $T$ is temperature, $p$ is 
pressure, and, following the literature, we use $s$, the specific entropy, 
given by $S/M$, where $S$ is the usual entropy and $M$ is mass.  Similarly, $u 
= U/M$ is specific internal energy, $\nu = V/M$ is the specific volume, and $g 
= u - Ts + p\nu$ is the specific Gibbs energy.  For clarity, subscripted 
brackets, $\{\ldots\}_{T,\ p}$, are used to indicate that both temperature and 
pressure are held constant.  We imagine a membrane as a two-dimensional fluid 
only one molecule thick.  The fluid is an incompressible mixture of rodlike 
amphiphiles (henceforth referred to simply as lipids) and pointlike ``other'' 
molecules.  The other molecules are pointlike in the sense that they have no 
orientation, or, more specifically, we associate with them no energy 
contribution that is a function of the membrane shape.  (They are still assumed 
to have mass and to occupy volume in the normal way).  For simplicity, we take 
a highly idealised approach and assume that all such pointlike molecules are 
thermodynamically equivalent (though it should be noted that a full 
multi-component treatment is still possible, if tedious).  In this simplified 
case, the specific Gibbs energy is just
\begin{equation}
	g = \sum_k \mu_k c_k + g^\mathrm{nem},
	\label{eq:g^nem}
\end{equation}
where, in the first term, the concentration and chemical potential of component 
\(k\) are given by \(c_k = M_k / M\) and \(\mu_k\), respectively, with 
\(k\in\{l,o\}\) (\(l\) for ``lipids'', and \(o\) for ``other'').  If the 
nonamphiphilic molecules were known, then the subscript would simply label 
them, in turn, \(k\in\{l,1,2,3,\ldots\textit{etc}.\}\).  The second term in 
(\ref{eq:g^nem}) is the energy associated with the nematic nature of the lipid 
molecules, which can be approximated by an elastic 
description~\cite{APIB84,SS+85} of the chemical interactions between the head 
groups and between the tails groups of the lipid molecules.  The rest of this 
section briefly reviews such a theory (and its assumptions) in order to 
highlight certain important caveats that arise due to our description being 
both dynamical and multicomponent.

\subsection{Elastic description of amphiphilic interactions}
The benefits of an elastic theory are twofold.  First, the system can be 
written formally in terms of a single well-defined ``neutral-torque'' surface 
\(\mathcal{S}\), using a standard procedure.  Second, it provides a description 
for the quantities $\kappa$, $\kappa_g$, and $C_0$ in terms of geometrical 
variables, allowing the assumptions of this approach to be made plain.  The 
standard route is to write \(g^\mathrm{nem}\) as a sum of two quadratic terms:
\begin{equation}
	g^\mathrm{nem} = k_\mathrm{h} \left( a_\mathrm{h} - 
	a_{\mathrm{h},0}\right)^2 + k_\mathrm{t} \left( a_\mathrm{t} - 
	a_{\mathrm{t},0}\right)^2,
	\label{eq:mono_2}
\end{equation}
where subscripts are used to indicate either head or tail groups.  For example, 
$a_\mathrm{h} = A_\mathrm{h} / M$ is the area per unit mass on a surface 
defined to intersect the head groups of all the molecules, while $a_\mathrm{t} 
= A_\mathrm{t} / M$ is defined in a similar way for the molecular tails.  Both 
$k_\mathrm{h}$ and $k_\mathrm{t}$ are constants, as are the quantities 
$a_{\mathrm{h},0}$ and $a_{\mathrm{t},0}$, defined as the area per unit mass at 
equilibrium for head and tail groups respectively.  The projection onto a 
neutral-torque surface is described in~\cite{EERS80,APIB84} and recapitulated 
(with some small modifications) in Appendix \ref{app:elastic}.  The result is 
that, apart from a term which describes the free energy associated with a 
lateral tension (which for our model is assumed constant) 
Eq.~(\ref{eq:Helfrich}) is recovered.  However, contrary to Helfrich's 
approach, the values of $\kappa$, $\kappa_g$, and $C_0$ are no longer constant, 
and depend on $a$, the local area per mass [see Eqs.~(\ref{eq:k}) to 
(\ref{eq:kappa_g})].  This conflict has previously been resolved by the authors 
of~\cite{LM+94} by minimising the free energy under the constraint of a fixed 
number of particles in the monolayer (the result is that $a$ is constant up to 
$\mathcal{O}\left(\delta^2\right)$, where $\delta$ is the length along the long 
axis of the lipid molecules).  However, since our description includes the 
behaviour of nonamphiphilic molecules, this result can only be applied as a 
lowest order approximation when the concentration of nonamphiphiles is low.  It 
is therefore important to state that, physically, we assume to always be in the 
regime of constant bending rigidity, when the nonamphiphilic molecules do not 
materially affect $\kappa$ and only contribute to free energy through the first 
term of (\ref{eq:g^nem}).

\subsection{Gaussian curvature}
It is also necessary to remark on the Gaussian curvature $K$.  While $K$ is 
routinely ignored in equilibrium studies of vesicles \cite{US96} due to the 
Gauss-Bonnet theorem, such an approach cannot be applied here as it relies on 
the integration of the membrane energy over a closed surface.  However, 
progress can be made by closely following nematohydrodynamics, where a standard 
assumption is that the dominant contributions to the free energy of a nematic 
fluid come from terms proportional to the director and its first spatial 
derivatives \cite{JLE61,FML68}.  In our case, since we may identify the 
director with the normal $\boldsymbol{n}$ of the surface $\mathcal{S}$, this 
amounts to ignoring all terms in (\ref{eq:Helfrich}) that contain higher 
spatial derivatives of $\boldsymbol{n}$.  Leaving the details to Appendix 
\ref{app:geom} (geometry will also be discussed in the next section) the 
important point is that, alongside the standard formula 
$H=\nabla\cdot\boldsymbol{n}/2$, the Gaussian curvature can be shown to be 
given by
\begin{equation}
	K = \frac{1}{2}\left[\left(2H\right)^2 
	+\boldsymbol{n}\cdot\left(\nabla^2\boldsymbol{n}\right)\right],
	\label{eq:Kofn_main}
\end{equation}
which involves second spatial derivatives of $\boldsymbol{n}$.  As a result, 
fully incorporating the effects of the Gaussian curvature is left for further 
work, and in this treatment we retain only the first term of 
(\ref{eq:Kofn_main}).
Taking the above into account, the lipid interactions only contribute to the 
Gibbs energy via the mean curvature:
\begin{equation}
	g^\mathrm{nem} = \frac{\kappa}{2}\left( 2H - C_0 \right)^2 + 
	\frac{\kappa_g}{2}\left(2H\right)^2.
	\label{eq:g^nem_2}
\end{equation}
For the purposes of thermodynamics, it is helpful to identify extensive and 
intensive contributions in Eq.~(\ref{eq:g^nem_2}).  Therefore, noting from 
(\ref{eq:kappa_g}) that $\kappa_g$ is linear in $\kappa$, and in order to 
conform with the conventions of nematohydrodynamics, we define the intensive 
variable $\kappa' = \kappa / \rho$, and the corresponding extensive variable 
\(\psi = \frac{\rho}{2}\left( 2H - C_0\right)^2 + 
\frac{\rho\kappa_g}{2\kappa}\left( 2H\right)^2\), where \(\rho = M /V\) is the 
usual mass density.
From here, it follows that
\begin{equation}
	g^\mathrm{nem} = \kappa'\psi,\ \ \mathrm{and}\ \ \{\mathrm{d}g\}_{T,\ p,\ 
	c_i} = \kappa'\mathrm{d}\psi,
	\label{eq:g^nem_final}
\end{equation}
which, when returning to the local equilibrium condition, gives
\begin{equation}
	T\mathrm{d}s = \mathrm{d}u + p \mathrm{d}\nu - \sum_{k} \mu_k \mathrm{d}c_k 
	- \kappa'\mathrm{d}\psi.
	\label{eq:Gibbs_nematic}
\end{equation}

\section{LNET for (curved) membranes}
The usual LNET approach is to manipulate the time derivative of 
(\ref{eq:g^nem_final}) through careful application of conservation laws and 
constitutive relations and then to compare the result with the equation for 
local entropy balance,
\begin{equation}
	\rho \frac{\mathrm{d} s}{\mathrm{d} t} = - 
	\nabla\cdot\boldsymbol{J}_\mathrm{s} + \sigma.
	\label{eq:local_ent_bal}
\end{equation}
(Here $\boldsymbol{J}_\mathrm{s}$ is the convective entropy flux, and $\sigma$ 
is the entropy production term.)  Under normal circumstances, this tactic then 
identifies the source term with a bilinear sum of the thermodynamic forces and 
fluxes, from which Onsager relations can be deduced.

Before proceeding however, recall that these relations are now defined on the 
2D surface \(\mathcal{S}\), which is assumed to be regular and parametrized by 
variables \(u\) and \(v\).  Any point on the surface is then defined by the 
position vector $\boldsymbol{r} = \boldsymbol{r}\left(u,v\right)$, and the 
tangent space at each point is spanned by the two vectors 
$\boldsymbol{r}_\alpha \equiv \partial \boldsymbol{r} / \partial q^\alpha$.  
Here $\alpha,\ \beta\ \in\ \{1,2\}$--- and similarly for all Greek indices--- 
where $q^1 = u$ and $q^2 = v$.  Using the shorthand notation $\partial_\alpha 
\equiv \partial / \partial q^\alpha$, the gradient operator becomes
\begin{equation}
	\nabla \equiv g^{\alpha\beta}\boldsymbol{r}_\beta \partial_\alpha,
	\label{eq:nabla}
\end{equation}
where $g_{\alpha\beta} = \boldsymbol{r}_\alpha \cdot \boldsymbol{r}_\beta$ is 
the metric tensor and $g^{\alpha\beta}$ its inverse.  For membranes with a 
finite thickness, one must typically compute corrections to 
diffuse processes that are caused by curvature~\cite{NO10}.  However, since the 
monolayer is assumed to be only one molecule thick, these corrections can be 
ignored and most of the traditional conservation laws (\textit{e.g.}, mass and 
internal energy) carry over without any modifications.  For a system at 
constant uniform temperature, the application of these laws (Appendix
\ref{app:conserv}) gives
\begin{equation}
		\rho T\frac{\mathrm{d} s}{\mathrm{d} t} = - 
		\mathsf{\Pi}\colon\left(\nabla\boldsymbol{v}\right)^\mathsf{T} + \sum_k 
		\mu_k \nabla\cdot\boldsymbol{J}_k - \rho\kappa'\frac{\text{d} 
	\psi}{\text{d} t},
	\label{eq:dGibbs_dt_nematic_2}
\end{equation}
where \(\boldsymbol{J}_k = \rho_k\left( \boldsymbol{v}_k - \boldsymbol{v} 
\right)\) is the local diffusion flow, and \(\mathsf{\Pi} = \mathsf{P} - 
p\mathsf{I}\) is the nonhydrostatic part of the pressure tensor.  (Here $\rho_k 
\equiv M_k / V$ are partial mass densities, and $\boldsymbol{v} \equiv \sum_k 
\boldsymbol{v}_k \rho_k / \rho$ defines both the barycentric velocity 
$\boldsymbol{v}$, and the partial velocities $\boldsymbol{v}_k$.)  The notation 
uses sans-serif font for a (rank 2) tensor, a superscript \(\mathsf{T}\) to 
denote the transpose, and a colon to represent the trace of an interior 
product; \textit{i.e.}, the term 
$\mathsf{\Pi}\colon\left(\nabla\boldsymbol{v}\right)^\mathsf{T}$ is written in 
component form as
$\Pi_{ij}g^{\alpha\beta}\left(r_\alpha\right)_i \partial_\beta v_j$, where a 
sum is implicit for repeated indices, both Greek and Latin.  Here Latin indices 
are the usual Cartesian components in three dimensions such that \( 
(r_\alpha)_i \) is the $i$-th component of the tangent vector 
\(\boldsymbol{r}_\alpha\).

As previously mentioned, the traditional liquid crystals approach for computing 
the time-derivative of free energies with nematic order--- such as the final 
term in (\ref{eq:dGibbs_dt_nematic_2})--- is to assume a dependence on 
$\boldsymbol{n}$ and its first spatial derivatives $\nabla\boldsymbol{n}$ and 
then impose straightforward constitutive relations for these 
quantities~\cite{JLE61,FML68}.  For our case, since $\psi$ was constructed with 
these constraints in mind, the final term in (\ref{eq:dGibbs_dt_nematic_2}) can 
be computed directly from \(H=\nabla\cdot\boldsymbol{n}/2\), although as we 
show, care must be taken when differentiating the gradient operator.  
Specifically, by recalling the definition of $\psi$, the time derivative can be 
written as
\begin{equation}
	\frac{\mathrm{d}\psi}{\mathrm{d} t} =
	\frac{\partial \psi}{\partial \rho}\frac{\mathrm{d}\rho}{\mathrm{d} t} + 
	\frac{\partial \psi}{\partial \left(\nabla\cdot\boldsymbol{n}\right)}
		\frac{\mathrm{d}\left(\nabla\cdot\boldsymbol{n}\right)}{\mathrm{d} t}.
		\label{eq:time_deriv}
\end{equation}
Here if the gradient that appears in the last term was just the usual operator 
(and not restricted to the surface), then
\begin{equation}
	\frac{\mathrm{d}}{\mathrm{d}t} \left( \nabla\cdot\boldsymbol{n} \right)
	= \nabla\cdot\frac{\mathrm{d}\boldsymbol{n}}{\mathrm{d}t},
	\label{eq:corrolary}
\end{equation}
and equations similar to those that appear in the theory of nematic liquid 
crystals could be recovered~\cite{JLE61,FML68}.  In turn, this leads to terms 
in (\ref{eq:dGibbs_dt_nematic_2}) that are either convective or bilinear in 
forces and fluxes, which is required by LNET and therefore desirable.  With 
this in mind, the part that is of interest in (\ref{eq:time_deriv}) is the 
proper time-derivative of $\nabla\cdot\boldsymbol{n}$, which in component form 
becomes
\begin{equation}
	\begin{split}
	\frac{\mathrm{d}}{\mathrm{d} t} \left[ 
		g^{\alpha\beta}\left(r_\alpha\right)_i \partial_\beta n_j\right] =& 
		\frac{\mathrm{d}}{\mathrm{d} t} \left[ 
			g^{\alpha\beta}\left(r_\alpha\right)_i \right]\partial_\beta n_j
		+\\& g^{\alpha\beta}\left(r_\alpha\right)_i 
		\frac{\mathrm{d}}{\mathrm{d} t} \left[\partial_\beta n_j\right].
			\label{eq:time_deriv_2}
	\end{split}
\end{equation}
At this stage we notice that the director $\boldsymbol{n}$ is an average over 
local orientations of the amphiphilic molecules only; therefore, we may use the 
material derivative
\begin{equation}
	\frac{\mathrm{d}}{\mathrm{d}t} = \frac{\partial}{\partial t} + 
	\boldsymbol{v}_l\cdot \nabla,
	\label{eq:hydro_d_dt}
\end{equation}
where the velocity is that of the lipids (amphiphiles).  Applying 
(\ref{eq:hydro_d_dt}), the last term in (\ref{eq:time_deriv_2}) can be expanded 
(again, in component form) to give
\begin{equation}
	\begin{split}
	g^{\alpha\beta}\left(r_\alpha\right)_i \frac{\mathrm{d}}{\mathrm{d} t} 
	\left[\partial_\beta n_j\right] = & 
	g^{\alpha\beta}\left(r_\alpha\right)_i\partial_\beta\left(\frac{\partial 
	n_j}{\partial t}\right) + \\&
	g^{\alpha\beta}\left(r_\alpha\right)_i g^{\mu\nu} 
	\left(v_l\right)_k\left(r_\mu\right)_k\partial_\nu\partial_\beta n_j.
\end{split}
	\label{eq:expand}
\end{equation}
Manipulating derivatives, one can then show that the second term of 
(\ref{eq:expand}) is just equivalent to
\begin{equation}
	g^{\alpha\beta}\left(r_\alpha\right)_i \partial_\beta \left[ g^{\mu\nu} 
	\left(v_l\right)_k\left(r_\mu\right)_k\right]\partial_\nu n_j,
	\label{eq:manip}
\end{equation}
which, when combined with (\ref{eq:time_deriv_2}) makes it clear that the 
condition (\ref{eq:corrolary}) is only satisfied if
\begin{equation}
	\frac{\mathrm{d}}{\mathrm{d}t} \left( 
	g^{\alpha\beta}\boldsymbol{r}_\alpha\right)
	=
	\nabla \left[ 
		g^{\alpha\beta}\left(
			\boldsymbol{v}_l\cdot\boldsymbol{r}_\alpha
		\right)
	\right].
	\label{eq:assumption}
\end{equation}
The simplest way of understanding the ramifications of this relation is to note 
that it causes the distortion stress to vanish (apart from a straightforward 
density-dependent term).  That is, it is implicitly assumed that there is no 
energy cost (and therefore no force per unit area) arising as a result of 
deformations which move the relative positions of the molecules but keep the 
director field fixed.  Mathematically, this is certainly self-consistent with a 
free energy that is only based on the relative orientation of the molecules, 
\textit{viz.} Eq.~(\ref{eq:mono_2}).  Physically, since both the lateral 
tension and the bending rigidity are constant and uniform, any membrane 
configuration that leads to the same director field must be thermodynamically 
equivalent.

The main corollary of (\ref{eq:assumption}) is that it implies
\begin{equation}
	\begin{split}
		\rho T\frac{\mathrm{d} s}{\mathrm{d} t} &= 
		-\nabla\cdot\boldsymbol{J}^\mathrm{nem} - 
		\mathsf{\Pi}'\colon\left(\nabla\boldsymbol{v}\right)^\mathsf{T}\\
		&\quad+\sum_k \mu_k \nabla\cdot\boldsymbol{J}_k - \boldsymbol{h}\cdot 
		\frac{\mathrm{d}\boldsymbol{n}}{\mathrm{d}t},
	\end{split}
	\label{eq:nematic_included}
\end{equation}
where
\begin{equation}
	\boldsymbol{h} = - \kappa'\nabla \left(\rho\frac{\partial\psi}{ \partial 
		\left(\nabla\cdot\boldsymbol{n}\right)}\right) = 
		-2\rho\bar{\kappa}\nabla H,
		\label{eq:h_alpha}
\end{equation}
is the molecular field (constant density), $\mathsf{\Pi}' = \mathsf{\Pi} - 
\kappa'\mathsf{I}\rho^2 \left(\partial\psi / \partial\rho\right)$ is the 
viscous stress, and $\boldsymbol{J}^\mathrm{nem}$ is a convective ``nematic 
flux'' term.  [We have introduced the shorthand $\bar{\kappa}=(\kappa + 
\kappa_g) / \rho$, for simplicty.] Before making the comparison with 
(\ref{eq:local_ent_bal}) however, a number of constraints can be implemented to 
make the analysis simpler.  First, it is standard to separate out the 
temperature dependence by manipulating derivatives.  Second, we must recognise 
that the barycentric diffusion flows are not independent; \textit{i.e.}, 
\(\boldsymbol{J}_l + \boldsymbol{J}_o = 0\).  Finally, since we stipulate that 
all molecules point normal to the surface, contributions from the antisymmetric 
part of \(\mathsf{\Pi}\) can be neglected (this arises from conservation of 
angular momentum; see~\cite{LiquidCrystalsDeGennes,NonEquilThermDeGrootMazur} 
and
Appendix \ref{app:ang_mom}).  With these modifications, one may identify the 
entropy production term as
\begin{equation}
	\begin{split}
		\sigma =& - \frac{1}{T}\mathsf{\Pi}^\mathsf{s} 
		\colon\left(\nabla\boldsymbol{v}\right)^\mathsf{s} - 
		\frac{1}{T}\boldsymbol{J}_l \cdot \left\{\nabla  \left(\mu_l - 
		\mu_o\right)\right\}_{T,\ p,\ \kappa'}\\& -\frac{1}{T}\boldsymbol{h}
		\cdot\frac{\mathrm{d}\boldsymbol{n}}{\mathrm{d}t},
	\end{split}
\label{eq:sigma_nematic_final}
\end{equation}
where a superscript $\mathsf{s}$ is used to indicate the symmetric part.  It is 
this form that will lead to reciprocal relations of the Onsager type.

\section{Reciprocal linear relations and cross-diffusion}
Invoking the Cure principle and only writing linear relations for forces and 
fluxes of the same tensorial character, we focus on the coupled vector 
relations,
\begin{equation}
	\begin{split}
		\boldsymbol{J}_l =& 
		-L_\mathrm{dd}\frac{\left\{\nabla\left(\mu_l-\mu_o\right)\right\}_{T,\ 
		p,\ \kappa'}}{T} - L_\mathrm{dn} \frac{\boldsymbol{h}}{T},\\
		\frac{\mathrm{d}\boldsymbol{n}}{\mathrm{d}t} = & 
		-L_\mathrm{nd}\frac{\left\{\nabla\left(\mu_l-\mu_o\right)\right\}_{T,\ 
		p,\ \kappa'}}{T} - L_\mathrm{nn} \frac{\boldsymbol{h}}{T},
	\end{split}
	\label{eq:Onsager_vector}
\end{equation}
where the Onsager coefficients $L_\mathrm{dd},L_\mathrm{nn}$ and 
$L_\mathrm{nd}=L_\mathrm{dn}$ are labeled by subscripts ``d'' for diffusion and 
``n'' for nematic.  It is these linear relations that couple lipid diffusion to 
curvature and which contain the main result of this paper.  However, the result 
is best cast in terms of a cross-diffusion effect such as the Soret or Dufour 
effects~\cite{NonEquilThermDeGrootMazur}.  To see this, we use the Gibbs-Duhem 
relation for our system
\begin{equation}
	c_l \left\{ \nabla \mu_l\right\}_{T,\ p,\ \kappa'} + c_o \left\{ \nabla 
	\mu_o\right\}_{T,\ p,\ \kappa'} = 0.
	\label{eq:G-D}
\end{equation}
From here, using the fact the \(c_o + c_l = 1\), it is clear that
\begin{equation}
	\left\{\nabla\left(\mu_l-\mu_o\right)\right\}_{T,\ p,\ \kappa'} = 
	\frac{\mu^c_{ll}}{c_o} \nabla c_l,
	\label{eq:grad_c_l}
\end{equation}
where the shorthand notation of~\cite{NonEquilThermDeGrootMazur} has been used: 
$\mu^c_{ll}\equiv\left(\partial \mu_l / \partial c_{l}\right)_{T,\ p,\ 
\kappa'}$.  On physical grounds, we expect that $\mu^c_{ll}\vert_{c_1=0}=0$, 
that is, the rate of change of the lipid chemical potential with respect to the 
concentration of lipids is zero when the system can no longer accommodate any 
more lipids.  We may also reasonably expect that $\mu^c_{ll}$ is a 
monotonically decreasing function of $c_l$.  With these constraints in mind, we 
make the simplest assumption possible: that $\mu^c_{ll} = A^c_{ll}(1-c_l) = 
A^c_{ll}c_o$, where $A^c_{ll}=\mathrm{constant}$.  Substituting this result, 
along with (\ref{eq:h_alpha}), into the linear relations 
(\ref{eq:Onsager_vector}) and then taking the divergence on both sides of the 
resulting expressions gives
\begin{equation}
	\begin{split}
	\frac{\mathrm{d} c_l}{\mathrm{d} t} =& D_\mathrm{dd} \nabla^2 c_l - 2 
	D_\mathrm{dn} \nabla^2 H,\\
	\frac{\mathrm{d} H}{\mathrm{d} t} =& 
	-\frac{A^c_{ll}}{2\bar{\kappa}}D_\mathrm{nd}\nabla^2 c_l + D_\mathrm{nn} 
	\nabla^2 H,
\end{split}
	\label{eq:cross_diff}
\end{equation}
where both (\ref{eq:corrolary}) and the condition for conservation of mass, 
$\rho\left(\mathrm{d} c_l / \mathrm{d} t\right) = 
-\nabla\cdot\boldsymbol{J}_l$, have been used, and $\nabla^2$ is the 
Laplace-Beltrami operator (Laplacian on the surface).  The new coefficients 
comprise two ``direct'' terms, a diffusion coefficient,
\begin{equation}
	D_\mathrm{dd} = \frac{L_\mathrm{dd} A^c_{ll} }{\rho T},
	\label{eq:D_dd}
\end{equation}
and a curvature relaxation coefficient,
\begin{equation}
	D_\mathrm{nn} = \frac{\rho\bar{\kappa}}{L_\mathrm{nn}}{T},
	\label{eq:D_nn}
\end{equation}
plus two indirect cross-terms,
\begin{equation}
	D_\mathrm{dn} = D_\mathrm{nd} = \frac{L_\mathrm{dn}\bar{\kappa}}{T},
	\label{eq:D_cross}
\end{equation}
that represent molecular diffusion induced by curvature and its reciprocal 
effect of curvature induced by molecular diffusion, respectively.

\section{Discussion}
In summary, Eqs.~(\ref{eq:cross_diff}) describe the relaxation of a 
two-component (amphiphilic and nonamphiphilic) fluid membrane towards 
equilibrium at constant temperature and constant lateral tension.  The result 
was obtained by combining nematohydrodynamics and geometry under the important 
assumption (\ref{eq:assumption}), which is equivalent to the statement that 
\begin{equation} \frac{\partial}{\partial
	t}\left( g^{\alpha\beta}\boldsymbol{r}_\alpha\right) =
	g^{\alpha\beta}\boldsymbol{r}_\alpha\cdot
	\left(\nabla\boldsymbol{v}_l\right).  \label{eq:alternative}
\end{equation}
\begin{figure}[t]
	\centering
	\includegraphics[scale = 0.7]{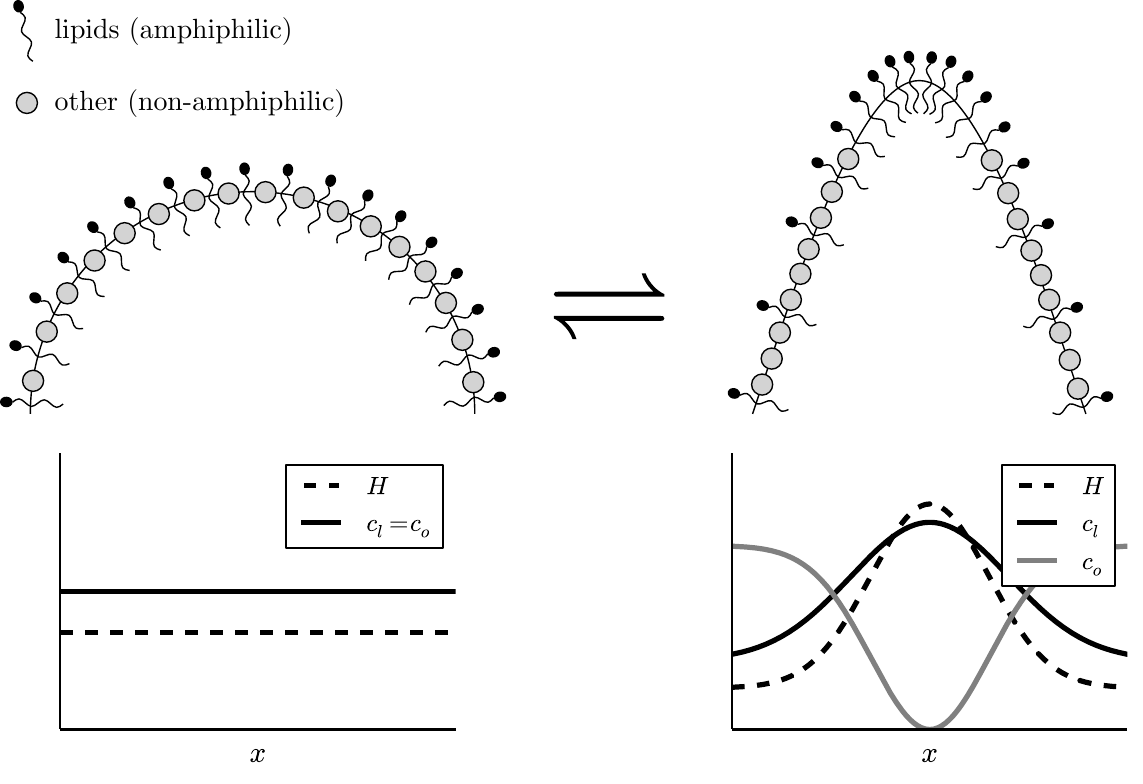}
	\caption
	{
		Exaggerated cartoon indicating the behaviour described by
		Eqs.~(\ref{eq:cross_diff}).  At constant curvature, the concentrations 
		of both amphiphilic and non-amphiphilic components are spatially 
		uniform.  Bending the membrane so that the curvature has a gradient 
		with non-zero divergence (\textit{e.g.}, parabolic, as shown) induces a 
		flow of amphiphilic molecules towards the region where $\nabla^2 H$ is 
		largest (\textit{i.e.}, the tip), displacing any nonamphiphilic 
		molecules.  The reverse process is also allowed: Starting with a 
		parabolic membrane and nonuniform concentrations, the system relaxes to 
		a state of both spatially uniform curvature and molecular 
		concentrations.
	}
\label{fig:schematic}
\end{figure}
That is, the explicit time dependence of the tangent vectors (and their 
weights) is fixed to be a function of the velocity gradients.  Or, more 
heuristically, since the lipids (amphiphiles) are constrained to point normal 
to the surface, there is no way of deforming the membrane without moving the 
relative positions of the amphiphiles and nonamphiphiles.  The result predicts 
that a membrane with a curvature that has nonzero Laplacian--- \textit{e.g.}, a 
parabolic profile--- will relax towards a constant-curvature surface 
accompanied by a simultaneous flow of lipids away from regions where the 
Laplacian of the curvature is largest (see Fig.~\ref{fig:schematic}).  Indeed, 
the equations also predict the inverse effect: If a concentration gradient 
(with nonzero divergence) can be established in a free membrane, then it would 
induce a corresponding curvature.  An intuitive understanding of this effect 
can be gained by thinking in terms of molecular splay: the degree to which the 
head groups of the lipids are separated with respect to their tails. In regions 
of high curvature, the splay--- and hence the energy associated with the 
configuration--- is very high.  However, since the molecules are fluid, they 
can address this situation by displacing nonamphiphilic molecules to increase 
their concentration--- and reduce their splay--- at points where the gradient 
of the curvature has non-zero divergence.  Similarly, an enforced (zero 
divergence) concentration gradient of lipids is not optimal on a 
constant-curvature surface due to the repulsive effect of too little splay 
[recall the quadratic profile (\ref{eq:mono_2})] and hence the surface--- free 
to move--- adjusts its conformation accordingly.

We remark that the form of diffusion coefficients (\ref{eq:D_dd}) to 
(\ref{eq:D_cross}) acts reinforce this picture.  For example, if the effective 
bending modulus $\bar{\kappa}$ is increased--- \textit{i.e.}, by using 
different lipids--- then the rate of the cross-diffusion is also increased.  
That is, if the membrane is stiffer, and more energy has to be added to the 
system to induce, say, a parabola (as in Fig.~\ref{fig:schematic}), then the 
rate at which lipids flow to reduce the energy of the system is increased by 
the same factor.

Such effects should, in principle, be observable by experiments that monitor 
single particle diffusion \cite{qdotsinmemb} on a pseudo-free membrane (very 
large vesicle or sheet with zero hydrostatic pressure difference), where 
numerical estimates for the diffusion constants, and hence Onsager coefficients 
would be welcome.  Indeed, the broader theory predicts other cross-effects, 
notably in the presence of a temperature gradient, although this is considered 
out of the scope of this study.

Finally, while there are undoubtedly drawbacks and limitations inherent with
such an approach, it is hoped that the work presented here can be adopted and
extended in order to overcome any difficulties.  In particular, a better
understanding of the types (and behaviour) of nonamphiphilic molecule most
prevalent in biological membranes would be helpful.  Armed with such 
information, it is plausible that this kind of model could shed light on the 
role of biological membranes in interesting but unexplained phenomena, such as 
cell locomotion.  We therefore welcome further work in the area.

\begin{acknowledgments}
R.~G.~M.~thanks the grant CEA$/$DSM-Energie for financial support when working 
at the Institut de Physique Th\'{e}orique (CEA).   Similarly, 
R.~G.~M.~acknowledges the current EPSRC grant EP/E501311/1 and also thanks 
A.~J.~McKane of the University of Manchester for providing the initial interest 
in the subject along with many helpful comments and suggestions at the outset.
\end{acknowledgments}

\appendix

\section{Elastic description of a monolayer}
\label{app:elastic}
In order to better understand the nematic contributions from the specific Gibbs 
energy, one may adopt an elastic theory of monolayers attributed to 
Ref.~\cite{SS+85} but also described in Ref.~\cite{APIB84}.  Retaining the 
spirit of the original work, we present a variant of this approach and 
demonstrate how it should be interpreted within the context of LNET.  Central 
to the approach is the asymmetric nature of amphiphilic molecules.  The 
principal idea is that, when close to equilibrium, the free energy of a 
monolayer can be modelled as two elastic sheets with different elastic moduli.  
The two elastic sheets approximate the different chemical interactions between 
the head groups and between the tail groups, of amphiphilic molecules.  As 
described in the main text, it is assumed that $g^\mathrm{nem}$ is a sum of two 
quadratic terms, repeated here for convenience:
\begin{equation}
	g^\mathrm{nem} = k_\mathrm{h} \left( a_\mathrm{h} - 
	a_{\mathrm{h},0}\right)^2 + k_\mathrm{t} \left( a_\mathrm{t} - 
	a_{\mathrm{t},0}\right)^2.
	\label{eq:mono_SI}
\end{equation}
Here, subscripts are used to indicate either head or tail groups.  For example, 
$a_\mathrm{h} = A_\mathrm{h} / M$ is the area per unit mass on a surface 
defined to intersect the head groups of all the molecules, whilst $a_\mathrm{t} 
= A_\mathrm{t} / M$ is defined in a similar way for the molecular tails.  Both 
$k_\mathrm{h}$ and $k_\mathrm{t}$ are constants, as are the quantities 
$a_{\mathrm{h},0}$ and $a_{\mathrm{t},0}$, defined as the area per unit mass at 
equilibrium for head and tail groups respectively.  The problem with 
incorporating the above into a two-dimensional thermodynamic description is 
that Eq.~(\ref{eq:mono_SI}) is technically defined on two separate surfaces.  
The rest of this appendix therefore follows the approach described by Evans and 
Skalak~\cite{EERS80} in order to write the Gibbs energy in terms of variables 
defined on a single common surface.

Consider a sample of monolayer under planar stress, that is, $a_\mathrm{h} = 
a_\mathrm{t} = a$.  The equilibrium separation of the molecules when confined 
to a plane is then given by the value of $a$ for which $g$ is a maximum.  More 
formally,
\begin{equation}
	\frac{\partial g^\mathrm{nem}}{\partial a} = 2k_\mathrm{h} \left( a - 
	a_{\mathrm{h},0}\right) + 2 k_\mathrm{t} \left( a - a_{\mathrm{t},0}\right) 
	= 0,
	\label{eq:mono_planar}
\end{equation}
which implies
\begin{equation}
	a = \frac{k_\mathrm{h} a_{\mathrm{h},0} + k_\mathrm{t} 
	a_{\mathrm{t},0}}{k_\mathrm{h} + k_\mathrm{t}} \equiv a_0.
	\label{eq:a_0}
\end{equation}
In order to achieve a planar strain on the monolayer it is necessary to apply 
an asymmetric stress.  For example, if $k_\mathrm{t}$ is bigger than 
$k_\mathrm{h}$ then the  tension applied in the same plane as the tails will 
have to be larger than that applied to the heads.  Mathematically, the lateral 
tension applied to the head groups while maintaining planar equilibrium is 
defined as
\begin{equation}
	\Omega_\mathrm{h} = \left.\frac{\partial g^\mathrm{nem}}{\partial 
		a_\mathrm{h}}\right\vert_{a_{\mathrm{h}}=a_0},
\end{equation}
and similarly for $\Omega_\mathrm{t}$.  It is then possible to define the 
so-called neutral surface, which exists between the head and tail groups and is 
uniquely defined as the surface of points for which the moments acting on the 
molecules in planar equilibrium are zero.  Here, the word planar is important, 
as true equilibrium would obviously lead to a curved monolayer (for 
$k_\mathrm{h} \neq k_\mathrm{t}$). The surface is defined by the equation
\begin{equation}
	\Delta \Omega_\mathrm{h} \delta_\mathrm{h} + \Delta \Omega_\mathrm{t} 
	\delta_\mathrm{t} = 0,
	\label{eq:neutral}
\end{equation}
where $\delta_\mathrm{h}$ and $\delta_\mathrm{t}$ define the distance from the 
neutral surface to the head and tail groups respectively.  The quantity $\Delta 
\Omega_\mathrm{h}$ is the small change in lateral tension acting at the head 
groups which arises from a small change in the planar separation.  From the 
definitions above, it is clear that $\Delta \Omega_\mathrm{h} = 2 k_\mathrm{h} 
\Delta a$ and similarly for $\Delta \Omega_\mathrm{t}$.  Substituting into 
Eq.~(\ref{eq:neutral}) it follows that
\begin{equation}
	\frac{\delta_\mathrm{h}}{\delta_\mathrm{t}} = 
	\frac{k_\mathrm{t}}{k_\mathrm{h}}.
	\label{eq:neutral_2}
\end{equation}
From here, it is then possible to find expressions for $a_\mathrm{h}$ and 
$a_\mathrm{t}$ in  terms of $a$, the area per molecule on the neutral surface, 
and distances $\delta_\mathrm{h}$ and $\delta_\mathrm{t}$.  Geometrical 
relations of this type are discussed in detail in~\cite{SchaumTenCalc}, 
therefore here it suffices to simply state the results
\begin{equation}
	a_\mathrm{h} = a\left[ 1 + \delta_\mathrm{h}2H + \delta_\mathrm{h}^2 K + 
		\mathcal{O}\left( \delta_\mathrm{h}^3 \right)  \right],
	\label{eq:a_h}
\end{equation}
and
\begin{equation}
	a_\mathrm{t} = a\left[ 1 - \delta_\mathrm{t}2H + \delta_\mathrm{t}^2 K + 
		\mathcal{O}\left( \delta_\mathrm{t}^3 \right)  \right],
	\label{eq:a_t}
\end{equation}
where two assumptions have been made: first, that $\delta_\mathrm{h}$ and 
$\delta_\mathrm{t}$ are of the same order of magnitude, and, second, that the 
thickness of the membrane is small on the scale of any reasonable curvature.   
As such, terms of order greater than $\delta_\mathrm{h}^2$ and 
$\delta_\mathrm{t}^2$  have been neglected.  Substituting Eqs.~(\ref{eq:a_h}) 
and (\ref{eq:a_t}) into Eq.~(\ref{eq:mono_SI})  leads to an expression for the 
free energy in terms $a$, $H$, and $K$.  Using Eq.~(\ref{eq:neutral_2}) and the 
fact that $\delta = \delta_\mathrm{h} + \delta_\mathrm{t}$, the resultant 
expression can be manipulated so that terms of the same order in $\delta$ may 
be grouped together.  This gives the expressions referenced in the main text,
\begin{equation}
	g^\mathrm{nem} = k\left( a - a_0 \right)^2 + \frac{\kappa}{2}\left( 2H - 
	C_0 \right)^2 + \kappa_g K,
	\label{eq:mono_3}
\end{equation}
where
\begin{equation}
	k = k_\mathrm{h} + k_\mathrm{t},
	\label{eq:k}
\end{equation}
\begin{equation}
	\kappa = \frac{2 k_\mathrm{h} k_\mathrm{t} a^2 \delta^2}{k},
	\label{eq:kappa}
\end{equation}
\begin{equation}
	C_0 = \frac{a_{\mathrm{h},0} - a_{\mathrm{t},0}}{a \delta},
	\label{eq:C_0}
\end{equation}
and
\begin{equation}
	\kappa_g = 2\kappa\left[ 1 - \frac{a_{\mathrm{h},0} + a_{\mathrm{t},0} - 
	a_0}{ka} \right].
	\label{eq:kappa_g}
\end{equation}

\section{Geometry}
\label{app:geom}
The two-dimensional surface (embedded in three dimensions) which represents the 
membrane is defined by a vector field $\boldsymbol{r} = \boldsymbol{r} (u,v)$, 
where $u$ and $v$ parametrize the surface. The tangent (vector) space 
associated with each point on the surface is then spanned by vectors 
$\boldsymbol{r}_\alpha \equiv \partial \boldsymbol{r} / \partial q^\alpha$, 
where $\alpha \in \{1, 2\},\:q^1 = u$ and $q^2 = v$.  From here, the first 
fundamental form, or metric, is defined as
\begin{equation}
	g_{\alpha\beta} \equiv \boldsymbol{r}_\alpha \cdot \boldsymbol{r}_\beta,
	\label{eq:metric}
\end{equation}
where the inverse metric $g^{\alpha\beta}$ is defined such that 
$g^{\alpha\beta}g_{\beta\gamma} = \delta^\alpha_\gamma$, with 
$\delta^\alpha_\gamma$ the Kronecker $\delta$ symbol.  Here, $g$ is the 
determinant of $g_{\alpha\beta}$, given by
\begin{equation}
	g \equiv 
	\frac{1}{2}\varepsilon^{\alpha\gamma}\varepsilon^{\beta\nu}g_{\alpha\beta}g_{\gamma\nu},
	\label{eq:g}
\end{equation}
where $\varepsilon^{\alpha\beta}$ is an antisymmetric two-dimensional 
Levi-Civita symbol.  The determinant is used to define the surface area 
element,
\begin{equation}
	\mathrm{d}A\equiv \sqrt{g}\mathrm{d}u\mathrm{d}v,
	\label{eq:dA}
\end{equation}
and the unit normal,
\begin{equation}
	\hat{\boldsymbol{n}} \equiv \frac{\boldsymbol{r}_1 \times 
	\boldsymbol{r}_2}{\sqrt{g}},
	\label{eq:normal}
\end{equation}
where, for notational simplicity, the traditional ``hat'' notation is omitted 
going forwards.  In order to quantify the curvature of a surface it is further 
necessary to define second derivatives $\boldsymbol{r}_{\alpha\beta} \equiv 
\partial^2\boldsymbol{r}/
\partial q^{\alpha}\partial q^{\beta}$, where the coefficients of the second 
fundamental form,
\begin{equation}
	L_{\alpha\beta} \equiv \boldsymbol{r}_{\alpha\beta} \cdot \boldsymbol{n},
	\label{eq:L_lm}
\end{equation}
and their determinant,
\begin{equation}
	l \equiv 
	\frac{1}{2}\varepsilon^{\alpha\gamma}\varepsilon^{\beta\nu}L_{\alpha\beta}L_{\gamma\nu},
	\label{eq:L}
\end{equation}
allow us to make contact with (\ref{eq:Helfrich}) by writing
\begin{equation}
	H \equiv -\frac{1}{2}g^{\alpha\beta}L_{\alpha\beta}
	\label{eq:H}
\end{equation}
and
\begin{equation}
	K \equiv \frac{l}{g},
	\label{eq:K}
\end{equation}
where a sum is implicit over repeated indices.  For consistency with 
(\ref{eq:Helfrich}) and the majority of membrane related literature, 
(\ref{eq:H}) is defined here contrary to the usual conventions of differential 
geometry, so that the mean curvature of a sphere is positive, 
$H_{\mathrm{sphere}} = 1 / R$.

In order to perform the derivatives present in the definition of the molecular 
field, it is necessary to write the mean and Gaussian curvatures as functions 
of the unit normal and its spatial derivatives.  We start with the mean 
curvature.  Since $\boldsymbol{r}_\alpha\cdot\boldsymbol{n} = 0$ then 
$L_{\alpha\beta} = -\boldsymbol{r}_\alpha\cdot\boldsymbol{n}_\beta$.  
Substituting into (\ref{eq:H}) and rewriting in terms of the gradient operator,
\begin{equation}
	\nabla\equiv g^{\alpha\beta} \boldsymbol{r}_\alpha \partial_\beta,
	\label{eq:grad}
\end{equation}
gives the standard result
\begin{equation}
	H = \frac{1}{2}\nabla\cdot\boldsymbol{n}.
	\label{eq:Hofn}
\end{equation}
Rewriting the Gaussian curvature is slightly more involved.  We use two 
relations without proof \cite{VesYangLiuXie}: the Weingarten relation,
\begin{equation}
	\boldsymbol{n}_\alpha = 
	-L_{\alpha\beta}g^{\beta\gamma}\boldsymbol{r}_\gamma,
	\label{eq:Weingarten}
\end{equation}
and the lesser known result,
\begin{equation}
	g^{\alpha\beta}L_{\beta\gamma}g^{\gamma\nu}L_{\nu\alpha} = 4H^2 - 2K.
	\label{eq:lesserknown}
\end{equation}
Our starting point comes from combining these two equations, where it is 
straightforward to see that
\begin{equation}
	\left(\boldsymbol{n}_\alpha\cdot\boldsymbol{n}_\beta\right)g^{\alpha\beta} 
	= 4H^2 - 2K.
	\label{eq:start}
\end{equation}
Here the left-hand side can be rewritten by using the fact that 
$\partial_\alpha \left( \vert \boldsymbol{n}\vert^2 \right) = 0$ and therefore 
$\boldsymbol{n}\cdot\boldsymbol{n}_\alpha = 0$, which implies that 
$\left(\boldsymbol{n}_\alpha\cdot\boldsymbol{n}_\beta\right)g^{\alpha\beta} = - 
\boldsymbol{n}\cdot\partial_\alpha\left(g^{\alpha\beta}\partial_\beta 
\boldsymbol{n}\right)$.  Introducing the Laplace-Beltrami operator,
\begin{equation}
	\nabla^2 \equiv \nabla\cdot\nabla = 
	\frac{1}{\sqrt{g}}\partial_\alpha\left(\sqrt{g} 
	g^{\alpha\beta}\partial_\beta\right),
	\label{eq:LaplaceBeltrami}
\end{equation}
it is relatively simple to show that
\begin{equation}
	\boldsymbol{n}\cdot\nabla^2\boldsymbol{n} = 
	\boldsymbol{n}\cdot\partial_\alpha\left(g^{\alpha\beta}\partial_\beta 
	\boldsymbol{n}\right),
	\label{eq:ndotdellsq}
\end{equation}
and hence Eq.~(\ref{eq:start}) can be inverted to show that
\begin{equation}
	K = \frac{1}{2}\left[\left(\nabla\cdot\boldsymbol{n}\right)^2 
		+\boldsymbol{n}\cdot\left(\nabla^2\boldsymbol{n}\right)\right].
	\label{eq:Kofn}
\end{equation}

\section{Application of thermodynamic conservation laws}
\label{app:conserv}
As described in the main text, the local Gibbs relation can be transformed by 
applying thermodynamic conservation laws.  Conservation laws of this type are 
commonplace in fluid dynamics and therefore simply stated here, with the full 
physical justification provided elsewhere (see 
\textit{e.g.},~\cite{NonEquilThermDeGrootMazur}).  We start with local 
conservation of mass which is unchanged by the inclusion of nematic media, 
giving
\begin{equation}
	\frac{\mathrm{d} c_k}{\mathrm{d} t} = - \frac{1}{\rho}\nabla \cdot 
	\boldsymbol{J}_k,
	\label{eq:localconservemass}
\end{equation}
where
\begin{equation}
	\boldsymbol{J}_k = \rho_k\left( \boldsymbol{v}_k - \boldsymbol{v} \right),
	\label{eq:diffflow}
\end{equation}
is the local diffusion flow.  Here, $\rho_k \equiv M_k / V$ are the partial 
mass densities, while the relation $\boldsymbol{v} = \sum_k \boldsymbol{v}_k 
\rho_k / \rho$ defines both the barycentric (or centre-of-mass) velocity 
$\boldsymbol{v}$, and the partial velocities $\boldsymbol{v}_k$.  In a similar 
way, the conservation of local heat $q$ is also unchanged by nematic effects, 
with the standard relation given by
\begin{equation}
	\frac{\mathrm{d} q}{\mathrm{d} t}= -\frac{1}{\rho} 
	\nabla\cdot\boldsymbol{J}_\mathrm{q},
	\label{eq:localconserveheat}
\end{equation}
where $\boldsymbol{J}_\mathrm{q}$ is the local heat flow.  Before writing down 
the conservation of internal energy however, note that a general analysis of 
the anisotropic term $\{\mathrm{d}g\}_{T,\ p,\ c_k}$ leads to a consideration 
of friction that would not otherwise be present for a simple fluid.  As a 
result, it is necessary to briefly review the general equation of motion for a 
fluid, given by
\begin{equation}
	\rho\frac{\mathrm{d} \boldsymbol{v}}{\mathrm{d} t} = 
	-\nabla\cdot\mathsf{P},
	\label{eq:EoM}
\end{equation}
where $\mathsf{P}$ is the stress tensor.  Assuming that the constituents of the 
fluid are inelastic it is natural to decompose the pressure tensor into a 
hydrostatic part $p$, and a tensor $\mathsf{\Pi}$, such that
\begin{equation}
	\mathsf{P} = p\mathsf{I} + \mathsf{\Pi},
	\label{eq:pressure}
\end{equation}
where $\mathsf{I}$ is the identity matrix.  In general, if the constituent 
particles are anisotropic, $\mathsf{\Pi}$ and therefore $\mathsf{P}$ are not 
symmetric.  Taking this into account, local conservation of internal energy 
becomes
\begin{equation}
	\frac{\mathrm{d} u}{\mathrm{d} t} = \frac{\mathrm{d} q}{\mathrm{d} t} - 
	\frac{1}{\rho}\mathsf{\Pi}\colon\left(\nabla\boldsymbol{v}\right)^\mathsf{T} 
	- \frac{\mathrm{d} \nu}{\mathrm{d} t},
	\label{eq:localenergyconserve_nematic}
\end{equation}
where the notation is described in the main text.  Using this result alongside 
conservation of heat and conservation of mass--- 
Eqs.~(\ref{eq:localconserveheat}) and (\ref{eq:localconservemass}) 
respectively--- it follows that
\begin{equation}
		\rho T\frac{\mathrm{d} s}{\mathrm{d} t} = 
		-\nabla\cdot\boldsymbol{J}_\mathrm{q} - 
		\mathsf{\Pi}\colon\left(\nabla\boldsymbol{v}\right)^\mathsf{T}
		+\sum_k \mu_k \nabla\cdot\boldsymbol{J}_k - \rho\left\{\frac{\text{d} 
	g}{\text{d} t}\right\}_{T,\ p,\ c_k},
	\label{eq:dGibbs_dt_nematic_SI}
\end{equation}
which is exactly the Eq.~(10) from the main text if the identifications (6), 
also in the main text, are made.

\section{Conservation of angular momentum}
\label{app:ang_mom}
The main idea of this appendix is to temporarily imagine that the lipid 
molecules were \textit{not} fixed to point normal to the membrane, and then 
implement conservation of angular momentum by following 
\cite{LiquidCrystalsDeGennes}.  Once conservation angular momentum has been 
imposed, it is then easier to ascertain the impact of enforcing the ``Helfrich 
condition'': that the director is always normal to a surface.  (Since the 
material contained in Ref.~\cite{LiquidCrystalsDeGennes} is presented there in 
a disparate way across a number of chapters, this appendix provides a 
systematic, if terse, formulation which has the benefit of being notationally 
consistent with the main text.)

First, it is necessary to recognise that the free energy of a nematic is 
unchanged if the both the molecular positions {\it and} the director are 
rotated by small amount.  More formally, consider the following deformations to 
position vector $\boldsymbol{r}$ and director $\boldsymbol{n}$, respectively:
\begin{equation}
	\delta \boldsymbol{r} = \boldsymbol{\omega} \times \boldsymbol{r},\ 
	\mathrm{and} \ \delta \boldsymbol{n} = \boldsymbol{\omega} \times 
	\boldsymbol{n}.  \label{eq:delta_r_delta_n}
\end{equation}
In component form, these relations become
\begin{equation}
	\nabla_i\left( \delta r \right)_j = \varepsilon_{ijl}\omega_l\ \mathrm{and} 
	\ \left( \delta n \right)_i = \varepsilon_{ijl}\omega_j n_l,
	\label{eq:delta_r_delta_n_comp}
\end{equation}
where indices $i$, $j$, and $l$ label Cartesian components ($k$ is reserved for 
labelling the components of the mixture) and $\varepsilon_{ijl}$ is the totally 
antisymmetric Levi-Civita symbol.  It can be seen that 
the total variation in the Gibbs energy due to small deformations is given by
\begin{equation}
	\rho\delta g = -\Pi^\mathrm{d}_{ij}\nabla_i\left( \delta r \right)_j + 
	h_i\left(\delta n \right)_i + \nabla_i \left( \rho \frac{\partial 
	g}{\partial\left(\nabla_i n_j \right)}\left(\delta n\right)_j \right).
	\label{eq:rho_delta_g}
\end{equation}
Substituting Eqs.~(\ref{eq:delta_r_delta_n_comp}) into the above and setting 
$\delta g = 0$ gives
\begin{equation}
	0=-\Pi^\mathrm{d}_{ij}\varepsilon_{ijl}\omega_l + 
	h_i\varepsilon_{ijl}\omega_j n_l + \nabla_i\left( \rho \frac{\partial 
	g}{\partial\left( \nabla_i n_j \right)}\varepsilon_{jlm}\omega_l n_m 
	\right),
	\label{eq:delta_g=0}
\end{equation}
where, by relabeling the indices of the second term, a common factor of 
(constant vector) $\omega_l$ may be removed.  Integrating this result over the 
entire volume, and using the divergence theorem \cite{Arfken}, leads to the 
relation
\begin{equation}
	\begin{split}
		0 &= -\int \Pi^\mathrm{d}_{ij}\varepsilon_{ijl}\mathrm{d}V + \int 
		\varepsilon_{ijl} n_i h_j \mathrm{d}V\\
		&\quad + \int \mathrm{d}A_i \left( \rho\frac{\partial g}{\partial\left( 
		\nabla_i n_j \right)} \right)\varepsilon_{mjl} n_m,
	\end{split}
	\label{eq:delta_g=0_int}
\end{equation}
where $\mathrm{d}A_i$ is the $i$th component of the surface area element 
$\mathrm{d}\boldsymbol{A}$.  With this relation in mind, it is necessary to 
temporarily turn attention to the nematic nature of the molecules.  Due to 
their rod-like form, a sample of nematic material must obey conservation of 
angular momentum.  The rate of change of total angular momentum is given by
\begin{equation}
	\frac{\mathrm{d} L}{\mathrm{d} t} = \frac{\mathrm{d}}{\mathrm{d} t}\int 
	\rho \left( \boldsymbol{r} \times \boldsymbol{v} \right)\mathrm{d} V.
	\label{eq:L_dot}
\end{equation}
The time derivative may be taken inside this integral by using Liebniz's rule,
\begin{equation}
	\frac{\mathrm{d} L}{\mathrm{d} t} = \int \frac{\partial}{\partial t} \Big[ 
		\rho\left(  \boldsymbol{r}\times\boldsymbol{v} \right) \Big] \mathrm{d} 
		V + \int \rho \left(\boldsymbol{r}\times\boldsymbol{v} 
		\right)\boldsymbol{v}\cdot\mathrm{d}\boldsymbol{A}.
	\label{eq:dL_dt}
\end{equation}
Manipulating derivatives and using conservation of mass it follows that
\begin{equation}
	\frac{\mathrm{d} L}{\mathrm{d} t} = \int\rho \frac{\mathrm{d}}{\mathrm{d} 
t}\left(\boldsymbol{r}\times\boldsymbol{v} \right)\mathrm{d}V = -\int 
\boldsymbol{r}\times\left( \nabla\cdot\mathsf{P} \right)\mathrm{d}V.
	\label{eq:dL_dt_2}
\end{equation}
Here, the final step comes from the equation of motion (\ref{eq:EoM}).  The 
resultant expression may then be integrated by parts.  In component form this 
gives
\begin{equation}
	\frac{\mathrm{d} L}{\mathrm{d} t} = -\int \varepsilon_{ijl} r_j 
	P_{ml}\mathrm{d}A_m + \int\varepsilon_{ijl} P_{jl}\mathrm{d}V,
	\label{eq:dL_dt_3}
\end{equation}
where the second term on the right-hand side can be simplified further by 
splitting the pressure tensor into hydrostatic and tensor parts, $P_{ij} = 
\Pi_{ij} + p\delta_{ij}$; {\it cf.}~Eq.(\ref{eq:pressure}).  The result is that
\begin{equation}
	\frac{\mathrm{d} L}{\mathrm{d} t} = -\int \varepsilon_{ijl} r_j 
	P_{ml}\mathrm{d}A_m + \int\varepsilon_{ijl} \Pi_{jl}\mathrm{d}V.
	\label{eq:dL_dt_3b}
\end{equation}
However, following de Gennes~\cite{LiquidCrystalsDeGennes}, the rate of change 
of angular momentum is also equal to the total torque due to external stresses 
acting at the boundary, given by
\begin{equation}
	-\int\boldsymbol{r}\times\left( \mathsf{P}\cdot\mathrm{d}\boldsymbol{A} 
	\right),
	\label{eq:external_stress}
\end{equation}
plus the total torque on the director at the boundary
\begin{equation}
	\int \boldsymbol{n}\times\left( \rho\frac{\partial g}{\partial 
		\left(\nabla\boldsymbol{n}\right)} \cdot\mathrm{d}\boldsymbol{A} 
		\right).
	\label{eq:torques_director}
\end{equation}
In component form, this can be written as
\begin{equation}
	\frac{\mathrm{d} L}{\mathrm{d} t} = \int\mathrm{d}A_m \varepsilon_{ijl} 
	\left[ n_j \left(\rho \frac{\partial g}{\partial \left( \nabla_m n_l 
	\right)}\right) - r_j P_{ml} \right].
	\label{eq:dL_dt_4}
\end{equation}
Comparison with Eq.~(\ref{eq:dL_dt_3b}) leads to the result
\begin{equation}
	\int \mathrm{d}V\varepsilon_{ijl}\Pi_{jl} = \int \mathrm{d}A_m 
	\varepsilon_{ijl} n_j \left(\rho \frac{\partial g}{\partial \left(\nabla_m 
	n_l\right)}\right).
	\label{eq:dL_dt_result}
\end{equation}
Finally, this may be combined with Eq.~(\ref{eq:delta_g=0_int}) to eliminate 
the surface integral on the right-hand side.  The result is that
\begin{equation}
	\varepsilon_{ijl}\Pi'_{jl} = -\varepsilon_{ijl} n_j h_l.
	\label{eq:ansymm_Pi}
\end{equation}
In order to understand this, it is useful to consider decomposing the viscous 
stress tensor into symmetric and anti symmetric parts.  Introducing 
superscripts $\mathsf{s}$ and $\mathsf{a}$ to denote \textit{symmetric} and 
\textit{antisymmetric}, respectively, gives
\begin{equation}
	\Pi'_{ij} = \Pi^\mathsf{a}_{ij} + \Pi^\mathsf{s}_{ij},
\end{equation}
where the elements of the anti-symmetric part are defined in the following way: 
\begin{equation}
\begin{split}
	\Pi^{\mathsf{a}}_{12} &= -\Pi^{\mathsf{a}}_{21} = 
	\frac{1}{2}\varepsilon_{3jl}\Pi'_{jl},\\
	\Pi^{\mathsf{a}}_{23} &= -\Pi^{\mathsf{a}}_{32} = 
	\frac{1}{2}\varepsilon_{1jl}\Pi'_{jl},\\
	\Pi^{\mathsf{a}}_{31} &= -\Pi^{\mathsf{a}}_{13} = 
	\frac{1}{2}\varepsilon_{2jl}\Pi'_{jl}.
\label{eq:antisymm_Pi}
\end{split}
\end{equation}
The viscous stress tensor arises in the expression for entropy production in an 
inner-product with the transpose of the velocity gradient tensor, 
$\left(\nabla\boldsymbol{v}\right)^\mathsf{T}$.   Indeed, it is also possible 
to split the velocity gradient tensor into symmetric and antisymmetric parts.  
In this way, the dyadic product splits into two separate dyadics between 
symmetric and antisymmetric parts
\begin{equation}
	\mathsf{\Pi}\colon\left(\nabla\boldsymbol{v}\right)^\mathsf{T} = 
	\mathsf{\Pi}^\mathsf{s}\colon\left(\nabla\boldsymbol{v}\right)^\mathsf{s} + 
	\mathsf{\Pi}^\mathsf{a}\colon\left(\nabla\boldsymbol{v}\right)^{\mathsf{T},\ 
	\mathsf{a}},
	\label{eq:sym_antisym}
\end{equation}
where the notation $\left(\ldots\right)^{\mathsf{T},\ \mathsf{a}}$ indicates 
the antisymmetric part of the transpose (which is equal to the transpose of the 
antisymmetric part).  This can be expressed more easily in component form
\begin{equation}
	\Pi_{ij}\nabla_i v_j = \Pi_{ij}^\mathrm{s}\left(\nabla_i 
	v_j\right)^\mathrm{s} + \Pi_{ij}^\mathrm{a}\left(\nabla_i 
	v_j\right)^\mathrm{a},
	\label{eq:sym_antisym_ij}
\end{equation}
where repeated indices imply a sum.  Here, as with Eqs.~(\ref{eq:antisymm_Pi}), 
the three independent parts of the anti-symmetric velocity gradient tensor may 
be linked to the components of the vector
\begin{equation}
	\omega_{i} = \frac{1}{2} \varepsilon_{ijk} \nabla_j v_k.
\end{equation}
Combining this with Eqs.~(\ref{eq:antisymm_Pi}), the dyadic between 
anti-symmetric parts which arises in Eq.~(\ref{eq:sym_antisym_ij}) may be 
re-written as
\begin{equation}
	\Pi^{\mathrm{a}}_{ij}\left(\nabla_i v_j\right)^{\mathrm{a}} = 
	-\frac{1}{4}\varepsilon_{ijl}\varepsilon_{imn}\Pi'_{jl}\nabla_m v_n .
\label{eq:anti_symm_dyadic}
\end{equation}
Here, it is possible to invoke Eq.~(\ref{eq:ansymm_Pi}), the result of both 
symmetry considerations and angular momentum conservation.  It follows that
\begin{equation}
	\mathsf{\Pi}^\mathsf{a}\colon\left(\nabla\boldsymbol{v}\right)^{\mathsf{T},\ 
	\mathsf{a}} = 
	\frac{1}{2}\boldsymbol{\omega}\cdot\left(\boldsymbol{n}\times\boldsymbol{h}\right),
\end{equation}
where $\boldsymbol{\omega} = \left( \nabla \times \boldsymbol{v} \right) / 2$ 
is recognised as the vorticity.  Using the properties of the scalar triple 
product, it it clear that
\begin{equation}
	\mathsf{\Pi}'\colon\nabla\boldsymbol{v} + 
	\boldsymbol{h}\cdot\dot{\boldsymbol{n}} = 
	\mathsf{\Pi}^\mathsf{s}\colon\left(\nabla\boldsymbol{v}\right)^\mathsf{s} + 
	\boldsymbol{h}\cdot\boldsymbol{N},
	\label{eq:hn=nN}
\end{equation}
where $\boldsymbol{N} = \dot{\boldsymbol{n}} - 
\left(\boldsymbol{\omega}\times\boldsymbol{n}\right)/2$ is the rate of change 
of the director relative to the background fluid.

At this stage we simply apply the ``Helfrich condition'' by assuming that the 
director corresponds to the normal of a regular two-dimensional surface, 
embedded in three dimensions.  Immediately, one can see that 
$\mathsf{\Pi}^\mathsf{a}\colon\left(\nabla\boldsymbol{v}\right)^{\mathsf{T},\ 
\mathsf{a}} = 0$ and therefore $\boldsymbol{N} = \dot{\boldsymbol{n}}$.  The 
result is that:
\begin{equation}
	\mathsf{\Pi}'\colon\nabla\boldsymbol{v} + 
	\boldsymbol{h}\cdot\dot{\boldsymbol{n}} = 
	\mathsf{\Pi}^\mathsf{s}\colon\left(\nabla\boldsymbol{v}\right)^\mathsf{s} + 
	\boldsymbol{h}\cdot\dot{\boldsymbol{n}},
	\label{eq:hn=nN_amend}
\end{equation}
which is used in the result (15) of the main text.

\end{document}